\documentclass[a4paper,structabstract]{aa}
\usepackage{aas_macros}
\usepackage{natbib}
\usepackage{txfonts}
\usepackage{array}
\usepackage{wasysym} 

\usepackage{graphicx}

\def\modif{}
\def\rmod{}

\def\wig#1{\mathrel{\hbox{\hbox to 0pt{%
          \lower.6ex\hbox{$\sim$}\hss}\raise.4ex\hbox{$#1$}}}}

\def\mjup{$\rm M_J$}
\def\rjup{$\rm R_J$}

\def\teff{T_{\rm eff}}

\def\tint{T_{\rm int}}
\def\tirr{T_{\rm irr}}
\def\teq{T_{\rm eq}}
\def\frad{F_{\rm rad}}

\def\Jv{J_{\rm v}}
\def\Hv{H_{\rm v}}
\def\Kv{K_{\rm v}}
\def\kapv{\kappa_{\rm v}}

\def\Ith{I_{\rm th}}
\def\Jth{J_{\rm th}}
\def\Hth{H_{\rm th}}
\def\Kth{K_{\rm th}}
\def\kapth{\kappa_{\rm th}}
\def\fKth{f_{K{\rm th}}}
\def\fHth{f_{H{\rm th}}}

\def\tauch{\tau_{\rm ch}}
\def\tht{\tilde\theta}

\def\qnab{q\nabla T}

\def\aap{{\it Astron. \& Astrophys}}

\def\disp{\displaystyle}

\begin{document}

\title{On the radiative equilibrium of irradiated planetary atmospheres}
\titlerunning{Irradiated planetary atmospheres}
\authorrunning{Guillot}

\author{Tristan Guillot
 }

\offprints{T. Guillot}

\institute{Universit\'e de Nice-Sophia Antipolis, Observatoire de la C\^ote d'Azur, CNRS UMR 6202, B.P. 4229, 06304 Nice Cedex 4, France\\
           \email{tristan.guillot@oca.eu}}

\date{Submitted to A\&A, October 2, 2009; Accepted: June 4, 2010}

\abstract
{The evolution of stars and planets is mostly controlled by the properties of their atmosphere. This is particularly true in the case of exoplanets close to their stars, for which one has to account both for an (often intense) irradiation flux, and from an intrinsic flux responsible for the progressive loss of the inner planetary heat.}
{The goals of the present work are to help understanding the coupling between radiative transfer and advection in exoplanetary atmospheres and to provide constraints on the temperatures of the deep atmospheres. This is crucial in assessing whether modifying assumed opacity sources and/or heat transport may explain the inflated sizes of a significant number of giant exoplanets found so far.}
{I use a simple analytical approach inspired by Eddington's approximation for stellar atmospheres to derive a relation between temperature and optical depth valid for plane-parallel static grey atmospheres which are both transporting an intrinsic heat flux and receiving an outer radiation flux. The model is parameterized as a function of mean visible and thermal opacities, respectively.
}
{The model is shown to reproduce relatively well temperature profiles obtained from more sophisticated radiative transfer calculations of exoplanetary atmospheres. It naturally explains why a temperature inversion (stratosphere) appears when the opacity in the optical becomes significant compared to that in the infrared. I further show that the mean equivalent flux (proportional to $T^4$) is conserved in the presence of horizontal advection on constant optical depth levels. This implies with these hypotheses that the deep atmospheric temperature used as outer boundary for the evolution models should be calculated from models pertaining to the entire planetary atmosphere, not from ones that are relevant to the day side or to the substellar point. In these conditions, present-day models yield deep temperatures that are $\sim 1000\,K$ too cold to explain the present size of planet HD~209458b. An tenfold increase in the infrared to visible opacity ratio would be required to slow the planetary cooling and contraction sufficiently to explain its size. However, the mean equivalent flux is not conserved anymore in the presence of opacity variations, or in the case of non-radiative vertical transport of energy: The presence of clouds on the night side or a downward transport of kinetic energy and its dissipation at deep levels would help making the deep atmosphere hotter and may explain the inflated sizes of giant exoplanets.}
{}


\keywords{extrasolar giant planets -- planet formation}

\maketitle
%

\section{Introduction}

Many decades before computers would bring the possibility to model in detail the evolution of stars, an analytical solution to the problem of radiative transfer applied to stellar atmospheres revolutionized the study of their structure and evolution. This solution due to the English physicist Sir Arthur Stanley Eddington states that in a static plane-parallel grey stellar atmosphere in local thermal equilibrium, there is a simple relation between the temperature of the atmosphere and the optical depth of the radiation that it emits  \citep{Eddington1916}. 

Nowadays, detailed computer simulations are available both to find solutions for the complex problem of radiative transfer in stars and planets, and to study their evolutions. In particular, since the discovery of exoplanets, models to predict or account for their size have relied on more or less detailed radiative transfer models or tables \citep[][to cite just a few]{Guillot96,Bodenheimer01,Burrows03,Baraffe03, Fortney07}. Radiative transfer models in planetary atmospheres have highlighted the possibility of a bifurcation of solutions, depending on the presence or absence of efficient absorbers such as titanium oxyde and vanadium oxyde \citep{Hubeny03}, leading in some cases to temperature inversions \citep{Burrows07, Fortney08}. Observations of primary and secondary transits at different wavelengths have even brought the possibility to test these models and led to the identification of key molecular species \citep[e.g.][]{Tinetti07,Barman08,MS09,Swain10}.

However, the problem is complex and cannot be fully grasped by radiative transfer models that remain largely one-dimensional: this is because first the stellar irradiation field on the planet is intrinsically inhomogeneous and second atmospheric dynamics plays a crucial role in the global energy balance. In light of this, \citet{SG02} predicted that close-in giant planets should be characterized by significant day-night photospheric temperature variations, with the possilibity of an asymetry in the light curve due to heat transport by zonal winds. Both were verified by observations in the infrared \citep{Harrington06, Knutson07}. Detailed calculations combining radiative transfer and atmospheric dynamics are now being performed \citep[e.g.][to cite just a few]{Cho08,Langton08,Showman09,Menou09,Dobbs-Dixon10}, but are intrinsically limited by available computing power. 

In parallel, it has been realized that a significant fraction of irradiated giant planets are too large compared to what standard evolution models predict \citep[e.g.][]{Bodenheimer01,GS02,BHBH07,Guillot08,BCB08,MFJ09}. The possibility to choose as outer boundary conditions of evolution models between various atmospheric models, from those calculated with a maximal irradiation flux at the substellar point to those calculated with an irradiation averaged over the entire planet \citep[e.g.][]{Burrows03,Baraffe03} has left some confusion as to whether atmospheric properties may account for the discrepancy.

These are strong motivations towards the derivation of a simple analytical model to capture the important physics of the problem. {\modif This approach was already taken by \citet{Hansen08} who focused his analysis on the observational consequences in terms of emission from the planet}. In what follows, {\modif I will be mostly interested in understanding how the atmospheric properties impact the thermal evolution of planets}. I first derive the temperature-optical depth relation valid for an atmosphere which is heated both from below (intrinsic heat) and above (stellar irradiation) and compare it to some detailed atmospheric calculations from the literature. The study takes advantage of the fact that close-in giant planets should generally possess an extended radiative zone \citep{Guillot96}, so that convection may be neglected, as a first step at least. I then study the consequences of the variation of irradiation and advection on the mean temperature of the deep atmosphere. The resulting analytical temperature profile is used as boundary condition of evolution models, in an attempt to explain the inflated size of planet HD~209458b. Limitations to the models are detailed in section~\ref{sec:limitations}.

\section{An analytic radiative equilibrium model for irradiated atmospheres}

\subsection{Setting}
\begin{figure}
\centerline{\resizebox{8cm}{!}{\includegraphics{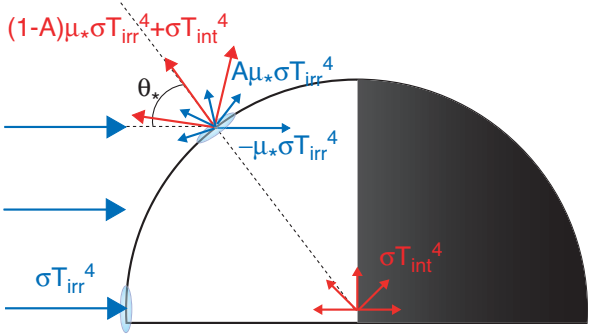}}}
\caption{Planet receiving a flux $\sigma\tirr^4$ from its parent star and emitting an intrinsic heat flux $\sigma\tint^4$. The labeled quantities correspond to radiative fluxes perpendicular to the atmospheric surface at the location considered. $\theta_*$ correspond to the angle between the direction of incidence of the collimated irradiation flux and the local vertical, and $\mu_*=\cos\theta_*$. Fluxes that are mostly characterized by visible wavelengths are drawn in blue. Fluxes in the infrared are drawn in red.}
\label{fig:scheme}
\end{figure}

The geometry of the problem is shown in fig.~\ref{fig:scheme}. The planet is receiving an irradiation flux $\sigma\tirr^4$, $\sigma$ being the Stefan-Boltzmann constant and $\tirr$ an effective temperature characterizing the irradiation intensity. In the cases to be considered, the incoming stellar irradiation can be considered as coming from a well-defined direction, with an angle $\theta_*$ to the perpendicular of the atmosphere. (The angular diameter of the irradiating star as seen from the atmosphere is $\alpha=\arctan(2*R_*/D)$, $R_*$ being the stellar radius and $D$ the star-planet distance. For an extremely close exoplanet $D\sim 5 R_*$ so that $\alpha\sim 22^\circ$, but in most cases of interest $\alpha<10^\circ$.) To first order, the irradiation temperature is a function of the stellar effective temperature $T_*$, its radius $R_*$ and the star-planet distance $D$:
\begin{equation}
\tirr=T_*\left(R_* \over D\right)^{1/2}.
\label{eq:tirr}
\end{equation}
At the substellar point on the planet, the flux received by the atmosphere is $\sigma\tirr^4$. On the day side of the planet, the flux perpendicular to the atmosphere is $\mu_*\sigma\tirr^4$, where $\mu_*\equiv\cos\theta_*$. (Note that eq.~(\ref{eq:tirr}) neglects the fact that the equator of the planet is slightly closer to the star than its poles, as this affects the irradiation temperature by a factor $\sim R_{\rm p}/2D$, $R_{\rm p}$ being the planetary radius, i.e. generally by $1\%$ or less.) It is useful to consider the irradiation flux averaged over the entire planetary surface, that I note $\sigma\teq^4$, following \citet{Saumon96}, and define the equilibrium temperature as:
\begin{equation}
\teq=T_*\left(R_* \over 2 D\right)^{1/2}.
\end{equation}
The maximal irradiation on the planet is $\sigma\tirr^4$ at the substellar point, the planet receives no flux on the night side, and the average irradiation is $\sigma\teq^4$, $1/4$th of the substellar point value. 

As indicated in fig.~\ref{fig:scheme}, at a given location in the atmosphere, a fraction $A\mu_*\sigma\tirr^4$ of the incoming flux is reflected, whereas $(1-A)\mu_*\sigma\tirr^4$ is absorbed in the atmosphere and eventually reemitted back to space. $A$ is the albedo, i.e. the fraction of the flux that is directly reflected. In general, this is a complex function of the properties of the atmosphere at the considered location, wavelength and direction. While the incoming flux is relatively highly collimated, this is not the case of the reflected, reemitted and intrinsic radiation fields. Furthermore, both the intrinsic and reemitted fluxes are mostly characterized by long wavelengths (in the infrared), with characteristic equivalent blackbody temperatures that are generally less than $2500$\,K. On the other hand, the irradiation and reflected fluxes are mostly characterized by short (optical) wavelengths, with equivalent blackbody temperatures equal to the effective temperature of the parent star, $\sim 5700$\,K for a solar-type star. As we seek an approximate solution to the radiative transfer problem, this will be important because it shows that the thermal and visible radiations are mostly decoupled. 

I now consider a given location in the atmosphere and a ray of intensity $I_{\nu\mu}$, $\nu$ being its frequency and $\theta$ its angle (or $\mu=\cos\theta$) with respect to the (local) vertical. Following \citet{Mihalas78} \citep[see also][]{Hubeny03}, I define three moments of the specific intensity as :
\begin{equation}
(J_\nu,H_\nu,K_\nu)\equiv {1\over 2}\int_{-1}^{1}I_{\nu\mu}(1,\mu,\mu^2)d\mu.
\end{equation}
$J_\nu$ is equivalent to the energy of the beam, $4\pi H_\nu$ is the radiation flux and $4 \pi/c K_\nu$ the radiation pressure. Given the parameters of the problem, $4\pi H_{\rm irr}=\int 4\pi H_\nu d\nu=\mu_* \sigma\tirr^4$ on the day side ($\mu_*>0$).

The {\rmod moments} of the radiative transfer equation in a static, plane-parallel atmosphere in local thermodynamic equilibrium and assuming isotropic scattering can be written \citep{Chandra60, Mihalas78}:
\begin{eqnarray}
&&  {\disp d H_\nu\over \disp  dm} = \kappa_\nu (J_\nu -B_\nu) \label{eq:H}\\
&&   { \disp d K_\nu\over \disp dm} = \chi_\nu H_\nu \label{eq:K}.
\end{eqnarray}
{\rmod Furthermore assuming radiative equilibrium implies:}
\begin{equation}
\int_0^\infty \kappa_\nu (J_\nu -B_\nu) d\nu =0 \label{eq:rad eq}
\end{equation}
where $m$ is the column mass ($dm=\rho dz$, $\rho$ being the density and $z$ the altitude), $\kappa_\nu$ is the true absorption coefficient, $\sigma_\nu$ the scattering coefficient and $\chi_\nu=\kappa_\nu+\sigma_\nu$ the total extinction coefficient. $B_\nu$ is the Planck function. {\rmod Using eq.~(\ref{eq:H}), the radiative equilibrium equation (\ref{eq:rad eq}) can be rewritten as a conservation equation for the total flux}, i.e.:
\begin{equation}
H\equiv\int_0^\infty H_\nu d\nu =\rm cte. \label{eq:Htot}
\end{equation}

A closure relation is needed to solve the problem. \citet{Eddington1916} noticed that in the interior of stars, one could consider as a first approximation $I_{\mu}=A+B\mu$, implying $J=A$ and $K=A/3$. Using $K/J=1/3$ in eq.~(\ref{eq:K}), and assuming a grey atmosphere $\kappa_\nu=\chi_\nu=\rm cte$ allowed solving the problem. 

Here I will follow the same approach but splitting the problem into two parts: one for the {\modif incoming radiation, mostly in the visible}, one for the {\modif outgoing radiation, mostly emitted in the infrared. Strictly speaking, the approach is valid in the limit when the incoming radiation and the outgoing radiation are always well separated in their characteristic wavelengths. One may then solve separately the radiation field of both radiation sources, determine the source function, and hence the interior temperature profile. In practice, this is only partially true for heavily irradiated exoplanets: as the temperature at depth increases, the thermal emission is pushed towards the visible. In some case, the photosphere may be so warm that the contribution in the visible is not negligible. In other cases, the irradiation flux may be characterized by low effective temperatures (e.g. if the parent star is an M-dwarf), so that the incoming and outgoing irradiation are not so different. However, in most cases we can consider that the region of the planetary atmosphere where the stellar irradiation is absorbed is characterized by temperatures that are significantly below the effective temperature of the star, so that the two radiation fields are mostly decoupled. This implies that characteristic mean opacities can be calculated for these two fields.}

\subsection{The incoming (visible) radiation}\label{sec:visible}

I first seek a solution of the radiative transfer problem for the {\modif incoming} radiation {\modif in the visible}. This is a long-standing problem in planetary atmospheres, one for which the scattering of the incoming light by atmospheric particules is crucial in determining the fraction of flux that is absorbed \citep[e.g.][]{Chandra60, Meador80, Toon89, Goody89}. For giant exoplanets orbiting close to solar-type stars (i.e. with orbital periods shorter than 10 days), the fraction of the irradiation that is reflected back is generally very low, of order 20\% or less, both from theoretical calculations \citep[e.g.][]{Sudarsky03, Hood08} and from observations \citep[e.g.][]{Rowe08, Snellen09,Alonso09}. For simplicity, I hence choose to neglect scattering: $\chi_\nu\approx \kappa_\nu$.

I integrate the moments of the radiation field in the visible:
\begin{equation}
(\Jv,\Hv,\Kv)\equiv \int_{\rm visible} (J_\nu,H_\nu,K_\nu) d\nu.
\end{equation}
Similarly, I define a mean opacity:
\begin{equation}
\kapv\equiv \Jv^{-1} \int_{\rm visible} \kappa_\nu J_\nu d\nu.
\end{equation}
It is interesting to note that since the visible radiation field is set by the stellar irradiation, $\kapv$ can be calculated {\em a priori}.

{\modif I now make an important simplification: I assume that the thermal emission from the atmosphere at visible wavelengths always has a negligible contribution to the global energy budget and that one can hence assume that} $B_\nu\sim 0$ for $\nu$ in the visible. The equations~(\ref{eq:H}) and (\ref{eq:K}) can hence be simplified by integrating over visible wavelengths:
\begin{eqnarray}
&&  {\disp d \Hv\over \disp  dm} = \kapv\Jv \label{eq:Hv}\\
&&   { \disp d \Kv \over \disp dm} = \kapv\Hv \label{eq:Kv}
\end{eqnarray}
{\modif This assumption is justified at low visible optical depth where clearly the incoming irradiation flux is intense and the contribution from the atmosphere is comparatively small: $J_\nu\gg B_\nu$ for $\nu$ in the visible and $\tau_{\rm v}\ll 1$. It can be questioned deeper down in the atmosphere where most of the incoming flux has been absorbed. This region is generally characterized by a thermal component of the radiation field that is much more intense than the visible part, so that for {\rmod large enough $\tau_{\rm v}$ values}, both $\Jth\gg \Jv$ and $\Jth \gg \int_{\rm visible} B_\nu d\nu$. This justifies neglecting the visible part of the source function in this region as well.}

Following Eddington, I write $\mu_*\equiv \sqrt{\Kv/\Jv}$. Note that this approach is valid in two extreme cases: for isotropic irradiation \citep[e.g.][]{Hubeny03}, for which $\mu_*=1/\sqrt{3}$, or in the case of collimated visible irradiation \citep[e.g.][]{Meador80}, in which case $\mu_*=cos\theta_*$. Equations~(\ref{eq:Hv}) and (\ref{eq:Kv}) then write:
\begin{equation}
{d^2(\Jv,\Hv)\over dm^2}={\kapv^2\over \mu_*^2}(\Jv,\Hv).
\end{equation}
Because {\modif the incoming radiation is rapidly absorbed, it is fine to assume that } both $\Jv$ and $\Hv$ vanish at great depths ($m\rightarrow\infty$), therefore
\begin{equation}
(\Jv(m),\Hv(m))=(\Jv(0),\Hv(0))e^{-\kapv m/\mu_*}.
\label{eq:Jv Hv}
\end{equation}
{\modif Again, it should be noted that $\Jv$ characterizes the intensity of the incoming radiation field, and that the thermal contribution from the atmosphere at visible wavelengths has been neglected.} 

Furthermore, eq.~(\ref{eq:Hv}) implies that:
\begin{equation}
\Hv(0)=-\mu_*\Jv(0). \label{eq:Hv0}
\end{equation}
In the case of an incoming radiation flux that is considered as fully isotropic, there is an inconsistency as we should have $\Hv(0)=-\Jv(0)/2$. This inconsistency is at the heart of the Eddington approximation however. It is due to the fact that the incoming irradiation flux cannot remain fully isotropic because of the larger absorption of grazing rays. Only an approximate solution can be found by neglecting the dependence on direction. In the collimated beam case, the solution is exact in the limit of no scattering.

\subsection{The outgoing (thermal) radiation}

Let us now consider the thermal part of the radiation field. As previously, we obtain average quantities by integration over thermal wavelengths: 
\begin{equation}
(B,\Jth,\Hth,\Kth)\equiv \int_{\rm thermal} (B_\nu, J_\nu, H_\nu, K_\nu) d\nu.
\end{equation}
\begin{equation}
\kapth\equiv \Jth^{-1} \int_{\rm thermal} \kappa_\nu J_\nu d\nu .
\end{equation}
As previously for the mean visible opacity, $\kapth$ is a function of temperature that can be calculated {\em a priori} {\rmod for a known outgoing radiation field}. 

The system of equations~(\ref{eq:H}) to (\ref{eq:rad eq}) is integrated over {\modif thermal} wavelengths with the hypothesis that {\modif $B^{-1}\int_{\rm thermal}\kappa_\nu B_\nu d\nu \approx \kapth$}: 
\begin{eqnarray}
&&  {\disp d \Hth\over \disp  dm} = \kapth (\Jth -B) \label{eq:Hth}\\
&&   { \disp d \Kth\over \disp dm} = \kapth \Hth \label{eq:Kth}\\
&&   \kapth (\Jth -B) +\kapv\Jv =0 \label{eq:rad eq th}
\end{eqnarray}

We can combine eqs.~ (\ref{eq:Jv Hv}), (\ref{eq:Hth}) and (\ref{eq:rad eq th}) to find by integration in $m$:
\begin{equation}
\Hth-\Hth(0)=-\mu_* \Jv(0)\left(1-e^{-\kapv m/\mu_*}\right). \label{eq:dHth}
\end{equation}
Separately, integrating eq.~(\ref{eq:Kth}) over column mass and using eq.~(\ref{eq:dHth}) yields:
\begin{equation}
\Kth-\Kth(0)=\int_0^m \kapth \left[ \Hth(0)-\mu_* \Jv(0)\left(1-e^{-\kapv m/\mu_*}\right)\right]dm.
\end{equation}
Equations~(\ref{eq:Htot}) and (\ref{eq:Hv0}) imply that $\Hth(0)=H+\mu_* \Jv(0)$, which allows integrating the above equation to find
\begin{equation}
\Kth-\Kth(0)=H\kapth m +\Jv(0) {\kapth \over \kapv} \mu_*^2 \left(1-e^{-\kapv m/\mu_*}\right).
\end{equation}

Using the first Eddington coefficient for the thermal radiation field, $\fKth\equiv \Kth/\Jth$, we obtain the mean intensity:
\begin{equation}
\Jth=\Jth(0)+H {\kapth m\over \fKth} +\Jv(0)  {\kapth \over \kapv} {\mu_*^2 \over \fKth} \left(1-e^{-\kapv m/\mu_*}\right) \label{eq:Jth final}
\end{equation}

\subsection{Atmospheric temperature profile}

\begin{figure}
  \centerline{\resizebox{\hsize}{!}{\includegraphics{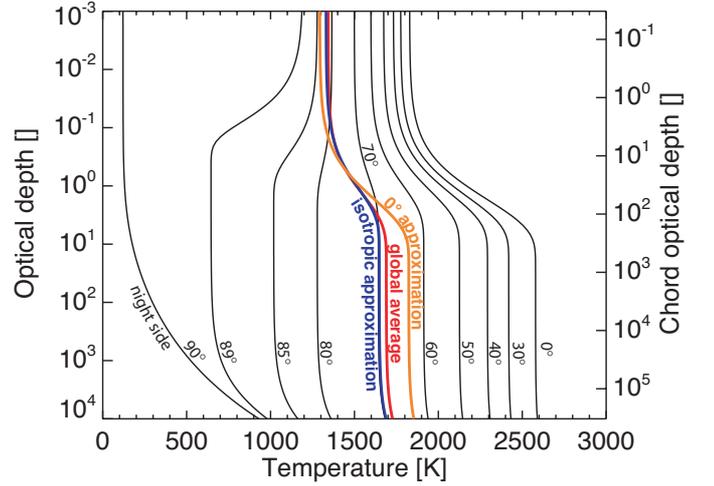}}}
  \caption{Temperature as a function of thermal optical depth obtained for different inclinations of the incident light, from (black lines, left to right)  $0^\circ$ (terminator) to $\theta_*=90^\circ$ (subsolar point) [see text and eq.~(\ref{eq:t4-mu})]. Also shown are the isotropic average [eq.~(\ref{eq:t4-isotropic})], $0^\circ$ average [eq.~(\ref{eq:t4-mu})] and global average [eq.~(\ref{eq:t4-global})] (see text). The quantities used for the plot correspond approximately to the case of HD209458b, i.e. $\teq=1469\,$K, $\tint=100\,$K, $\kapth=10^{-2}\rm\,cm^2\,g^{-1}$, $\kapv=4\times 10^{-3}\rm\,cm^2\,g^{-1}$. The corresponding chord optical depth is indicated on the right axis [see eq.~(\ref{eq:tauchord}) with $H=550\,$km, $R=94370\,$km and $\tht=1$]. }
  \label{fig:compare_mus_tau}
\end{figure}

I now turn to the derivation of the atmospheric temperature profile. I first introduce a second Eddington coefficient to relate the thermal radiative flux to the mean thermal intensity at the outer boundary: $\fHth\equiv \Hth(0)/\Jth(0)$. I also define the optical depth 
\begin{equation}
\tau\equiv \kapth m
\end{equation}
and, following \citet{Hansen08}
\begin{equation} 
\gamma\equiv \kapv/\kapth.
\end{equation}
Using these relations, eqs.~(\ref{eq:Hv0}), (\ref{eq:rad eq th}) and (\ref{eq:Jth final}), we can express the source function at each level:
\begin{eqnarray}
B&=&H\left[{1\over \fHth} + {\tau\over \fKth}\right] \nonumber \\
&& - \Hv(0) \left[{1\over \fHth}+  {\mu_*\over\gamma \fKth}+\left({\gamma\over \mu_*}-{ \mu_*\over \gamma\fKth}\right) e^{-\gamma\tau/\mu_*}\right]. \label{eq:B}
\end{eqnarray}
The fluxes $H$ and $\Hv(0)$ correspond to the imposed heat fluxes at the bottom and top of the atmosphere, respectively (note that $\Hv(0)<0$, as it corresponds to an inward flux). 

In the case of the collimated incoming irradiation, $H=\sigma \tint^4/4\pi$; $\Hv(0)=-\mu_*\sigma \tirr^4/4\pi$. 
As previously for the visible flux, $\fKth=1/3$ is valid both for a collimated and an isotropic radiation field. The second Eddington coefficient is generally chosen to be either $\fHth=1/2$ or $\fHth=1/\sqrt{3}$. The former value is derived from the assumption of isotropy of the outgoing radiation field, while the latter can be shown to result from an isotropic scattering \citep{Chandra60, Mihalas78}. The $1/2$ value however yields a temperature that is closer to the exact solution at great depth. Using thus $\fKth=1/3$ and $\fHth=1/2$ yields:
\begin{eqnarray}
T^4&=&{3\tint^4\over 4}\left[{2\over 3}+\tau\right]\nonumber\\
&&+{3\tirr^4\over 4}\mu_*\left[{2\over 3}+  {\mu_*\over \gamma}+\left({\gamma\over 3\mu_*}-{\mu_*\over\gamma}\right) e^{-\gamma\tau/\mu_*}\right].
\label{eq:t4-mu}
\end{eqnarray}
{\modif The solution is equivalent to that obtained by \citet{Hansen08}, except for a different boundary closure relation, and the fact that eq.~(\ref{eq:t4-mu}) accounts for atmospheric heating due to visible radiation (see Appendix).}
The $\tau=0$ limit for the temperature is
\begin{equation}
T(\tau=0)=\left\{ {1\over 2} \tint^4 + {1\over 2} \tirr^4\mu_* \left(1+{\gamma\over {\rmod 2}\mu_*}\right)\right\}^{1/4}
\end{equation}
Thus, the temperature at the top of the atmosphere is lowest and equal to that of a non-irradiated atmosphere emitting a total flux $\sigma(\tint^4+\mu_*\tirr^4)$ only in the case when $\gamma\longrightarrow 0$. In all other cases, the absorption of part of the visible irradiation flux at high levels in the atmosphere pushes the temperature up there. In the limit of high values of $\gamma$, the temperature-pressure gradient can even become negative, which is observed in models in the case of strong TiO/VO absorption \citep{Hubeny03,Fortney08,Burrows08}. {\modif (Note that however in this case, non-grey effects may dominate and should be considered)}.

\begin{figure*}[btp]
  \sidecaption
  \includegraphics[width=12cm,clip]{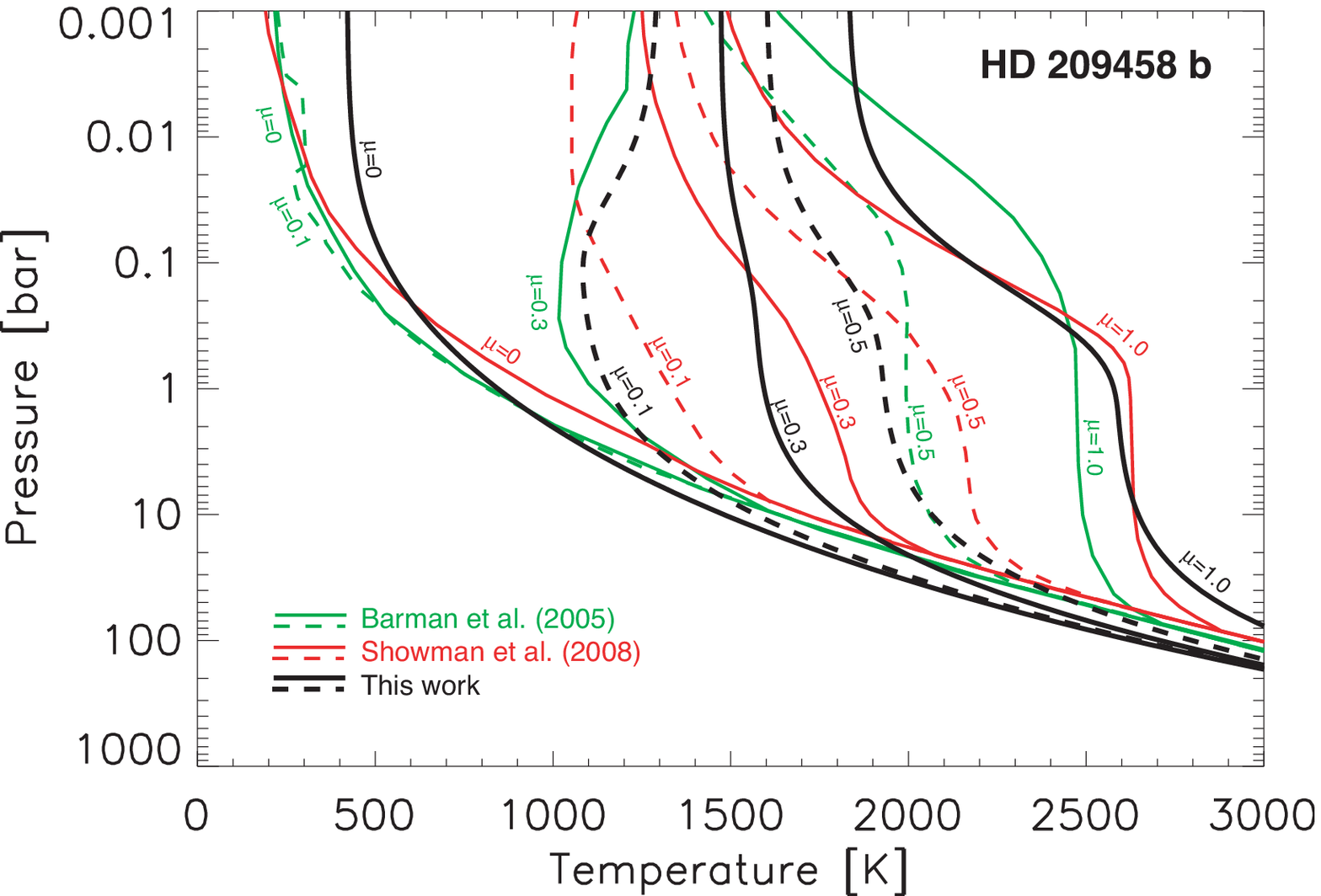}
 \caption{Atmospheric temperature-pressure profile obtained for different irradiation angles ($\mu=\cos\theta_*$) and compared to calculations for HD209458b available in the literature: \citet{Barman05} (in green -who assume $\tint=500\,$K on the night side to $\tint=230\,$K at the substellar point), and \cite{Showman08} (in red). The black lines are calculated as in fig.~\ref{fig:compare_mus_tau}, except that $\tint=500\,$K. The corresponding $\tau=2/3$ optical depth corresponds to the $P=65$\,mbar pressure level.}
  \label{fig:compare_mus}
\end{figure*}

 When assuming isotropic irradiation, the incoming flux is written $\Hv(0)=f\sigma \tirr^4/4\pi$, with $f=1$ at the substellar point, $f=1/2$ for a day-side average and $f=1/4$ for an averaging over the whole planetary surface \citep{Burrows03}. Furthermore, we have seen in \S\ref{sec:visible} that in this case $\mu_*=1/\sqrt{3}$, therefore:
\begin{eqnarray}
T^4&=&{3\tint^4\over 4}\left[{2\over 3}+\tau\right]\nonumber\\
&&+{3\tirr^4\over 4}f\left[{2\over 3}+  {1\over\gamma\sqrt{3}}+\left({\gamma\over \sqrt{3}}-{1\over\gamma \sqrt{3}}\right) e^{-\gamma\tau\sqrt{3}}\right].
\label{eq:t4-isotropic}
\end{eqnarray}
Of course, the standard Eddington relation is recovered in the limit when $\tirr\ll\tint$. 
{\modif For $\kapv=\kapth$, one gets that $T^4=3\tint^4/4(\tau+2/3)+0.93 f\tirr$, which is equivalent to the expression provided by \citet{Hubeny03} in that case}. 

Figure~\ref{fig:compare_mus_tau} provides a comparison of the temperature structures obtained for different values of the incident inclination $\theta_*$ and in the isotropic approximation, assuming an incoming flux that is averaged over the entire planetary surface ($f=1/4$). The numerical values have been chosen as representative of planet HD 209458b. Without advection and assuming a very small intrinsic heat flux, the temperature profile is found to vary dramatically between the substellar point ($\theta_*=0^\circ$) and the night side of the planet. The temperature becomes mostly isothermal at levels for which stellar irradiation has been entirely absorbed and before the contribution of the intrinsic heat starts to become significant. At higher levels, horizontal temperature variations on the day side are smaller because grazing rays are absorbed efficiently. For grazing incidences (or equivalently near the terminator), eq.~(\ref{eq:t4-mu}) always predicts a temperature inversion. Alternatively, lowering the ratio $\kapth/\kapv$ favors the formation of a temperature inversion even for vertical incidence. 

The comparison between the isotropic approximation and a $0^\circ$ approximation (i.e. assuming the atmosphere is at the substellar point but receives the irradiation flux that is the average for the entire planet) shows that the isotropic approximation yields deep temperatures that are smaller and temperatures at small optical depths that are larger. This is a direct consequence of the fact that grazing rays are absorbed at low optical depths.

\subsection{Comparison to models}

Dedicated temperature profiles at different incidences have been calculated for the atmosphere of HD209458b and are compared to the results of this work in fig.~\ref{fig:compare_mus}. The pressure was calculated by assuming constant gravity $g$, yielding $P=\tau g/\kapth$. Opacities in the visible and infrared were adjusted to obtain a good match to the more detailed models of \citet{Fortney08} and \citet{Showman08} at vertical incidence. As a result, this yielded $\kapth=10^{-2}\rm\,g\,cm^{-2}$ and $\kapv=4\times 10^{-3}\rm\,g\,cm^{-2}$. The figure shows that the match remains good for other incidences, and that differences between the analytic approximation and other models are of the same order as differences between the models themselves. Note that the model by \citet{Barman05} is a good match to the other solutions except at grazing incidences ($\mu=0.1$, $0.3$) where the temperature is found to be much lower. One likely possibility is that the plane-parallel approximation used both in this work and by \citeauthor{Fortney08} is overestimating the absorption at grazing incidences when compared to the more realistic spherical-symmetry approximation used by \citeauthor{Barman05}



\begin{figure}
\centerline{\resizebox{8cm}{!}{\includegraphics{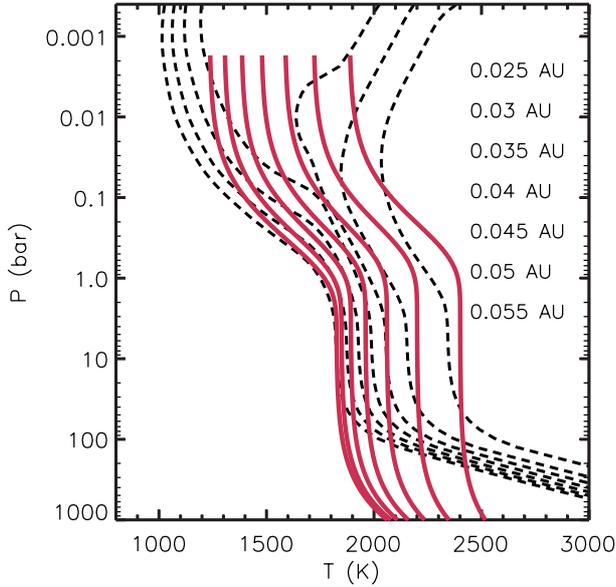}}}
\caption{Temperature profiles as a function of pressure obtained for different irradiation levels (corresponding to orbital distances between 0.025 and 0.055 AU) {\rmod at vertical incidence ($\mu_*=1$), using eq.~(\ref{eq:t4-mu}) (plain, red lines) and comparison to the results obtained by \citet{Fortney08} (dashed lines).}}
\label{fig:fortney}
\end{figure}

I now compare in fig.~\ref{fig:fortney} the solutions obtained from eq. ~(\ref{eq:t4-mu}) in the case of vertical incidence ($\mu_*=1$) to a detailed calculation by \citet{Fortney08} for planets at semi-major axes between 0.025 and 0.055\,AU from their star (assumed to be a solar twin). The planets have a $1\,$\mjup\ mass and a $1.2\,$\rjup\ radius, a solar-composition atmosphere, and the irradiation flux is calculated as a mean on the day-side hemisphere only. Again, the values of the opacity coefficients were adjusted to obtain a fair match, which was obtained for $\kapth=10^{-2}\,\rm cm^2\,g^{-1}$ and $\kapv=6\times 10^{-3} \sqrt{\tirr/2000\rm\, K}\,\rm cm^2\,g^{-1}$. These coefficients are indeed representative of values of the Planck or Rosseland means in these atmospheres, as is the weak temperature dependance on the opacity in the visible. They are compatible with the coefficients obtained specifically for the case of HD~209458b. The comparison also shows the limit of the model: at low pressures, the visible opacity is probably overestimated (except in the cases where TiO/VO are present), and at large pressures, the opacities are definitely underestimated. This is because collision-induced absorption and/or a rising electron abundance eventually pop in so that the mean opacities are to first order proportional to pressure (instead of being roughly independent of pressure).

\begin{figure}
\centerline{\resizebox{8cm}{!}{\includegraphics{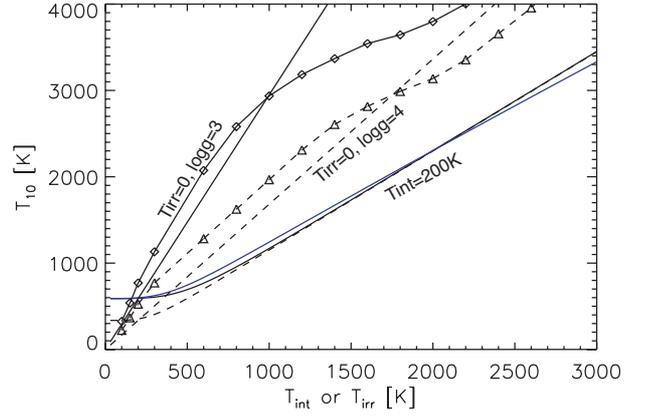}}}
\caption{Relation between intrinsic or irradiation effective temperature and the temperature at the 10\,bar level in planetary atmospheres. The line-styles indicate gravity: plain ($g=10^3\,\rm cm\,s^{-2}$) and dashed ($g=10^4\,\rm cm\,s^{-2}$). I assumed $\kappa_{\rm IR}=10^{-2}\,\rm cm^2\,g^{-1}$ and $\kappa_{\rm V}=6\times 10^{-3}\,\rm cm^2\,g^{-1}$, except for the thin blue line in the $\tint=200\,$K case and variable $\tirr$, for which $\kappa_{\rm V}=6\times 10^{-3}(T/2000\rm K)^{1/2}\,\rm cm^2\,g^{-1}$. The results of the analytical model for $\tirr=0$ are also compared to those calculated for a solar-composition atmospheres of isolated planets ($g=10^3\,\rm cm\,s^{-2}$, diamonds) and brown dwarfs ($g=10^4\,\rm cm\,s^{-2}$, triangles) by \citet{Saumon96}.}
\label{fig:tet10}
\end{figure}

The relation with the same opacity coefficients can then be usefully compared to a solar-composition model for isolated planets/brown dwarfs by \citet{Saumon96}. In this case, the 10 bar level is a useful value to tie the atmosphere (characterized by the part that is still at a relatively low optical depth) and the interior. This 10 bar level is also found to be within the isothermal layer for highly irradiated planets, so that it serves as a convenient outer boundary for the evolution models. Figure~\ref{fig:tet10} provides the value of $T_{10}$ as a function of $\tint$ or $\tirr$, in different cases, for the same values of $\kappa_{\rm IR}$ and $\kappa_{\rm V}$ as previously. For the non-irradiated cases ($\tirr=0$), one obtains a relatively fair match to the \citet{Saumon96} results. The model however separates from these numerical calculations both for temperature significantly lower or significantly higher than the usual $\sim 1000\,$K representative of giant planets at distances $\sim 0.1$\,AU and less to their star. This is due to changes in the mean opacities for these characteristic temperatures.

Figure~\ref{fig:tet10} also shows that in the case of a significant irradiation, the temperature is independent of gravity, and that $T_{\rm 10}\propto \tirr$. With eq.~(\ref{eq:t4-mu}) and $\mu_*=1$, this is easily explained by taking the limit $\tint\longrightarrow 0$ and $\tau\longrightarrow\infty$. In that case:
\begin{equation}
{T_{\rm deep}\over \tirr} = \left\{{3\over 4}\left({1\over\gamma}+{2\over 3}\right)\right\}^{1/4}
\end{equation}
With the fiducial opacity coefficients for HD209458b $\gamma=0.4$, I find that the right hand side is equal to $1.24$, close to the empirical $T_{10}/\tirr=1.25$ used by \citet{Guillot08} on the basis of models of irradiated planets calculated by \citet{Iro05}. This proportionality relation was also shown to apply to strongly irradiated atmospheres by \citet{Hubeny03}, who estimated $T_{10}/\tirr\approx 1.15$. 

\section{The non-uniform irradiation flux and the mean atmospheric temperature}

\subsection{Consequences of advection}

I now turn to the problem of the non-uniform irradiation flux, and its consequence for the deep atmospheric temperature that controls the global planetary evolution \citep{GS02,Arras06,Fortney07}. As we have seen, the solution to the pure radiative transfer problem is a temperature field that is intrinsically inhomogeneous. For giant planets in our Solar System, the combination of rapid rotation, long radiative timescales and of an intrinsic flux that is of the same order of magnitude as the incoming heat flux yields a relatively homogeneous temperature field \citep[e.g.][]{Ingersoll78}. On the contrary, close-in giant exoplanets should be locked in synchronous rotation \citep{Guillot96} and they are characterized by photospheric radiative timescales that are much shorter and intrinsic heat fluxes that are up to 4 orders of magnitude smaller than the irradiation fluxes \citep{GS02}. The question of the proper outer boundary condition to be used for evolution models is crucial. Planetary evolution models have been calculated either with boundary conditions inferred from calculations applying to the whole atmosphere, to the day-side hemisphere or even to the substellar point \citep[see][]{Burrows03, Baraffe03}. On the other hand, little attention has been paid to the consequences of the temperature inhomogeneities on the planetary cooling and of the validity of the different calculations. 

In \citet{GS02}, the problem of the planetary evolution with a non-uniform outer boundary condition had been approached by assuming that temperature differences may persist even deep down in the planet. In that case, an effective energy transport from the day side to the night side\footnote{Note that I use ``day side'' and ``night side'' for simplicity, but there is a strong variation in irradiation when moving from equator to pole that should be considered as well.} takes place simply to homogeneize the specific entropies at deep levels. As one would expect, the non-uniform outer boundary is found to lead to a faster cooling and contraction. 

However, the time-dependent radiative transfer models by \citet{Iro05} show that the radiative timescales increase very rapidly with depth, so that any remaining non-synchronous rotation or slow advection is susceptible to provide a very homogeneous temperature structure at depth. In this work, I will assume that there is a level, deep enough in the atmosphere/interior at which the temperature is independent of latitude/longitude. (Note that this should be deeper, peharps considerably, than the level at which the irradiation flux has been completely absorbed). I hereafter turn to the derivation of the temperature profile in an atmosphere that advects heat horizontally.

\subsection{Radiative transfer solution with advection}

I now consider that for each atmospheric location $(\theta,\phi)$ defined from the substellar point, mixing tables place by horizontal advection and transports heat with a flux $\qnab$. The radiative equilibrium equation becomes:
\begin{equation} 
\int_0^\infty \kappa (J_\nu -B_\nu) d\nu =\qnab
\end{equation}
or,
\begin{equation}
\kapth(\Jth-B)+\kapv\Jv=\qnab.\label{eq:B-q}
\end{equation}

The first moment of the radiative transfer equation (eq.~\ref{eq:H}) becomes by integration
\begin{equation}
\int_0^\infty {dH_\nu\over dm}d\nu=\qnab, 
\end{equation}
and hence
\begin{equation}
H(m)=H(\infty)-\int_m^\infty \qnab dm.
\end{equation}
Note that since we envision that $\nabla T\rightarrow 0$ when $m\rightarrow\infty$, this implies $H(\infty)=\sigma\tint^4/(4\pi)$. 

Now, the equation for $\Hth$ becomes:
\begin{equation}
{d\Hth\over dm}=-\kapv\Jv +\qnab
\end{equation}
and by integration
\begin{equation}
\Hth-\Hth(0)=-\mu_* \Jv(0)\left(1-e^{-\kapv m/\mu_*}\right)+\int_0^m \qnab dm'.\label{eq:Hth-q}
\end{equation}
$\Hth(0)=H(0)-\Hv(0)$ and therefore
\begin{equation}
\Hth(0)=H(\infty)-\int_0^\infty \qnab dm' + \mu_* \Jv(0).
\end{equation}
Inserting this relation into eq.~(\ref{eq:Hth-q}) yields
\begin{equation}
\Hth=H(\infty)+\mu_*\Jv(0) e^{-\kapv m/\mu_*} -\int_m^\infty \qnab dm'
\end{equation}

We now integrate the equation for the second moment of the radiation field:
\begin{eqnarray}
\Kth-\Kth(0)&=&H(\infty)\kapth m +\Jv(0){\kapth\over\kapv}\mu_*^2 \left(1-e^{-\kapv m/\mu_*}\right) \nonumber\\
&& -\int_0^m \kapth\int_{m'}^\infty \qnab dm'' dm'
\end{eqnarray}
and by integrating by parts:
\begin{eqnarray}
\Kth&=&\Kth(0)+H(\infty)\kapth m +\Jv(0){\kapth\over\kapv}\mu_*^2 \left(1-e^{-\kapv m/\mu_*}\right) \nonumber\\
&& -\kapth m \int_{0}^\infty \qnab dm' - \kapth \int_0^m (m'-m) \qnab dm'
\end{eqnarray}

The relation for $\Jth$ can then be found simply from the first Eddington coefficient $\fKth=\Kth/\Jth$. Then, using eq.~(\ref{eq:B-q}) yields
\begin{eqnarray}
B&=&\Jth(0)+H(\infty){\kapth m\over\fKth}+\Jv(0){\kapth\over\kapv}{\mu_*^2\over \fKth}\left(1-e^{-\kapv m/\mu_*}\right)\nonumber\\
&& -\kapth m \int_{0}^\infty \qnab dm' - \kapth \int_0^m (m'-m) \qnab dm' \nonumber\\
&& +\Jv(0){\kapv\over\kapth}e^{-\kapv m/\mu_*} -\qnab
\end{eqnarray}

We use the relations $\fHth\equiv \Hth(0)/\Jth(0)$, $\Hv(0)=-\mu_*\Jv(0)$,
\begin{equation}
\Hth(0)=H(\infty)-\Hv(0)-\int_0^\infty \qnab dm,
\end{equation}
and $H(\infty)=\sigma\tint^4$, $\Hv(0)=\mu_*\sigma\tirr^4$ to find an expression for the temperature profile at each location $(\tau,\mu,\phi)$ in the atmosphere:
\begin{eqnarray}
T^4&=&{3\tint^4\over 4}\left[{1\over 3\fHth}+{\tau\over 3\fKth}\right]\nonumber\\
&&+{3\tirr^4\over 4}\mu_*\left[{1\over 3\fHth}+  {\mu_*\over 3\gamma\fHth}+\left({\gamma\over 3\mu_*}-{\mu_*\over 3\gamma\fHth}\right) e^{-\gamma\tau/\mu_*}\right]\nonumber\\
&&-{\pi\over \sigma}\left\{\left({1\over \fHth}+{\tau\over\fKth}\right)\int_0^\infty \qnab dm\right. \nonumber\\
&&\hphantom{-{\pi\over \sigma}\left\{\right.}\left.+{\tau\over \fKth}\int_0^m\left({m'\over m}-1\right)\qnab dm' -\qnab\right\}
\label{eq:T4-adv}
\end{eqnarray}
The relation is a complex one and its resolution goes beyond the scope of the present article. 

\subsection{Mean atmospheric temperature}\label{sec:mean T}
We are mostly interested in the deep atmospheric temperature. As discussed, in the presence of an efficient-enough advection process, the temperature at deep levels should become latitudinally and longitudinally homogeneous. I therefore average over latitudes and longitudes (defined from the substellar point) to obtain a global mean temperature that depends only on depth $\tau$:
\begin{equation}
\overline{T^4}\equiv \oint T^4 d\omega ={1\over 4\pi}\int_0^{2\pi}\int_{-1}^1 T^4 d\mu d\phi.
\end{equation}
For a conservative advection scheme (in particular if $q$ does not depend on $\mu$, $\phi$ or $T$), $\oint\qnab d\omega=0$. {\rmod (One could easily show that this would also be true of heat diffusion, as long as the heat flux is conserved horizontally.)} This leads to a great simplification of eq.~(\ref{eq:T4-adv}) which becomes after integration over all latitudes and longitudes (using $\mu_*=\mu$):
\begin{eqnarray}
\overline{T^4}&=&{\tint^4\over 4}\left[{1\over \fHth}+{\tau\over \fKth}\right]+{1\over 2}{\tirr^4\over 4}Y\nonumber\\
Y&=&\int_0^1\mu\left[{1\over \fHth}+{\mu\over\gamma \fKth}+\left({\gamma\over \mu}-{\mu\over \gamma\fKth}\right) e^{-\gamma\tau/\mu}\right] d\mu.
\end{eqnarray}
Note that we integrated the intrinsic flux over the entire planet, whereas the irradiation flux is of course integrated only over the dayside hemisphere. 

The integral term can be rewritten
\begin{equation}
Y={1\over 2\fHth}+{1\over 3\gamma\fKth}+\gamma\int_1^\infty {e^{-\gamma\tau t}\over t^2}dt-{1\over \gamma\fKth}\int_1^\infty {e^{-\gamma\tau t}\over t^4}dt,
\end{equation}
or, in terms of exponential integrals $E_n(z)\equiv\int_1^\infty t^{-n}e^{-zt}dt$,
\begin{equation}
Y={1\over 2\fHth}+{1\over 3\gamma\fKth}+\gamma E_2(\gamma\tau)-{1\over \gamma\fKth}E_4(\gamma\tau).
\end{equation}
The $E_n$ functions have a recursive property \citep{abramowitz+stegun}:
\[ E_{n+1}(z)={1\over n}\left[e^{-z}-zE_n(z)\right]\]
which implies that with some algebra, $Y$ can be written more explicitly:
\begin{eqnarray}
Y&=&{1\over 2\fHth}+{1\over 3\gamma\fKth}\left[1+\left({\gamma\tau\over 2}-1\right)e^{-\gamma\tau}\right]\nonumber\\
&&+\gamma\left(1-{\tau^2\over 6\fKth}\right)E_2(\gamma\tau)
\end{eqnarray}

With our choice of $\fKth=1/3$ and $\fHth=1/2$, the equation for the mean temperature becomes
\begin{eqnarray}
\overline{T^4}&=&{3\tint^4\over 4}\left\{{2\over 3}+\tau\right\}+{3\teq^4\over 4}\left\{{2\over 3}+\right.\nonumber\\
&&\left. {2\over 3\gamma}\left[1+\left({\gamma\tau\over 2}-1\right)e^{-\gamma\tau}\right]+
{2\gamma\over 3}\left(1-{\tau^2\over 2}\right)E_2(\gamma\tau)\right\}
\label{eq:t4-global}
\end{eqnarray}
For $\gamma\tau\gg 1$, the properties of the $E_2$ function imply that
\[ \overline{T^4}\longrightarrow{3\tint^4\over 4}\left[{2\over 3}+\tau\right]
+{3\teq^4\over 4}\left[{2\over 3}+{2\over 3\gamma}\right]. \]
This relation is very similar to what obtained in the isotropic irradiation case but with a $(2/3\gamma)$ instead of a $(\sqrt{3})/\gamma$ coefficient. 

This is the same equation as obtained for the day side when considering no advection, but replacing $\tirr^4$ by $\teq^4$. There is hence a well defined mean temperature of the atmosphere at each level $\tau$ that is {\em independent} of the advective process to transport heat from the day side to the night side, as long as this advective process is conservative and takes place horizontally, over surfaces of constant optical depth $\tau$. 

Deep in the atmosphere ($\tau\gg 1$), the radiative time scale becomes very long, almost proportional to $P^2$ \citep{Iro05}. This means that any slow advection and/or any slightly asynchronous rotation would ensure an almost homogeneous mixing. Ideally, the location where this homogeneous mixing occurs should be used as an outer boundary condition for the interior and evolution models. Under these assumptions, only solutions in which the deep atmospheric temperature has been estimated with an irradiation flux averaged over the {\em entire} planet are energetically consistent.

\subsection{Comparison to models}\label{sec:comparison}

\begin{figure*}
\sidecaption
\includegraphics[width=12cm]{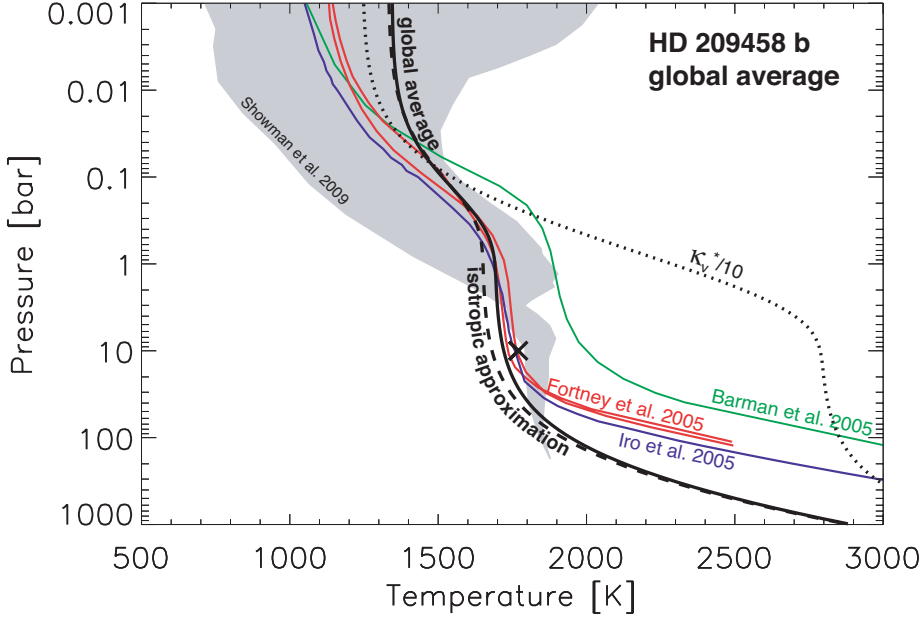}
\caption{Globally averaged atmospheric temperature-pressure profiles for HD~209458b. The plain black line corresponds to the result of a global average over all values of $\mu$, including the night side [eq.~(\ref{eq:t4-global})]. The dashed line is a result of the isotropic approximation for an irradiation flux averaged over the entire planetary surface [eq.~(\ref{eq:t4-isotropic})]. These are compared to similar calculations by \citet{Fortney05}, \citet{Iro05}, and \citet{Barman05}. The shaded region indicates the envelope of all possible temperature profiles found by \citet{Showman09}. The cross corresponds to the outer boundary condition used by \citet{Guillot06} and \citet{Guillot08} to calculate the evolution of the planet. The dotted line represents the globally averaged temperature-pressure profile obtained for a visible opacity reduced by a factor 10 compared to our fiducial value, and is representative of the atmospheric conditions that would allow explaining the evolution of HD209458b with no extra source of heat (see text). The parameters used for the calculation are as in fig.~\ref{fig:compare_mus_tau}, except I used $\tint=300$\,K. }
\label{fig:compare_4pi}
\end{figure*}

Figure~\ref{fig:compare_4pi} compares temperature profiles obtained for HD~209458b and relevant to the planet as a whole by various sources. First, as also shown in fig.~\ref{fig:compare_mus_tau}, it can be noticed that the global average calculation described by eq.~(\ref{eq:t4-global}) yields very slightly higher temperatures at depth, but is otherwise extremely close to the isotropic approximation of eq.~(\ref{eq:t4-isotropic}). The analytical solution (using the same values of the opacity coefficients as previously) is a good match to the more elaborate calculations by \citet{Fortney05} and \citet{Iro05}. These are however about 200 to 300\,K cooler than the temperature profile obtained by \citet{Barman05}. The figure also shows the envelope of solutions obtained in a dynamical circulation by \citet{Showman09} which is helpful to show the range of variability of the temperature profile in one particular model including radiative transfer and dynamical circulation. 

The cross in fig.~\ref{fig:compare_4pi} corresponds to the atmospheric boundary condition used in calculations of the evolution of HD209458b by \citet{Guillot08}. These yield a fast contraction of the planet and a radius that is at least 10\% smaller than the measurements \citep[e.g.][]{Knutson07}. The dotted line indicates the temperature profile obtained for a 10 times smaller visible opacity: because of a more efficient penetration of the incoming stellar flux, the deep levels are hotter than in the standard case by $\sim 1000\,K$. This is approximately the amount that is needed to account for the observed radius without invoking extra heat sources (see next section). As can be seen from the comparison of published radiative transfer calculations, this is outside the presently measured range of temperatures that is obtained by detailed models. The possibility that the visible opacity may be lowered that much is unlikely because even in the absence of efficient absorbers, scattering will have a non-negligible contribution. An alternative possibility is however to {\em increase} the infrared opacity by the addition of greenhouse gases at high altitudes while keeping the visible opacity to its nominal value, thus yielding a small $\gamma$ value. 

\begin{figure*}
  \sidecaption
  \includegraphics[width=12cm]{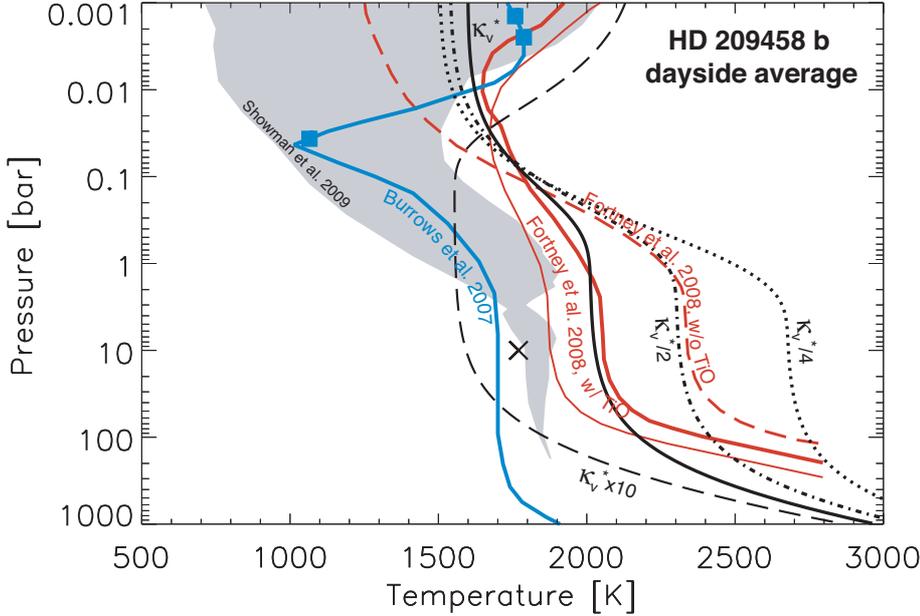}
  \caption{Atmospheric temperature-pressure profiles for HD~209458b averaged over the dayside hemisphere. The plain black line corresponds to a day-side average from eq.~(\ref{eq:t4-global}), with fiducial values for the opacities (see fig.1). Profiles resulting from visible opacities that are increased ten fold, decreased by 1/2 and 1/4 are also indicated by dashed, dash-dotted and dotted lines, respectively. Calculations for the dayside from \citet{Burrows07} are shown in blue, with squares indicating the inferred photospheric depth of Spitzer secondary eclipse measurements in 3 IRAC bands between 4.5 and 8 microns. Similar calculations from \citet{Fortney08} are indicated in red: three calculations are shown, with two models of TiO clouds, and without TiO and VO absorption, respectively. The envelope envelope of all possible temperature profiles found by \citet{Showman09} is indicated as a shaded region. The parameters used for the calculation are as in fig.~\ref{fig:compare_4pi}.}
  \label{fig:compare_2pi}
\end{figure*}

I now turn to calculations focused on characterizing the day-side hemisphere of HD209458b, motivated by secondary eclipse measurements. These calculations, in which the incoming stellar flux was averaged over the day-side hemisphere only are compared in fig.~\ref{fig:compare_2pi}. Particularly interesting are the calculations by \citet{Burrows07} and \citet{Fortney07} with TiO and VO clouds that have a temperature inversion and were found to be compatible with the colors measured with Spitzer during the secondary eclipse of the planet.  On the contrary, other temperature profiles with no inversion (e.g. the \citet{Fortney07} model with no TiO and VO clouds) were found to be incompatible. When applied to a day-side average, the analytic approximation is found to be a good match to the \citet{Fortney07} model with the same values of the opacities coefficients as previously. The \citet{Burrows07} model is colder by at least 300\,K, and can be more or less approximated by the analytical model with an order of magnitude increase of the visible opacity -thus yielding a pronounced temperature inversion-.

It should be noticed that the temperature inversion that appears required by observations is yielding lower deep temperatures. This is naturally explained by the fact that part of the incoming stellar flux is absorbed at greater levels and thus does not penetrate deep into the planet. One possibility to be investigated and that is out of the scope of the simple grey models presented here would be for the presence of significant non-grey absorbers: with opacities that are strongly wavelength-dependent, the energy in the center of absorption lines is absorbed high in the atmospheres, but the lower absorption in the wings allows for the possibility of a deeper penetration of energy at those wavelengths. In any case, fig.~\ref{fig:compare_2pi} highlights the fact that the present observational constraints relate to relatively high atmospheric levels, not to the deeper levels used to tie the atmosphere and the interior models.

\section{Atmospheric properties and the sizes of exoplanets}

I now reexamine the problem of the sizes of exoplanets in light of this atmospheric boundary condition with two parameters, $\kapth$ and $\kapv$. As before, HD~209458b is our proxy. First I rederive the difference in radius between the model radius, the photospheric radius and the transit radius. The problem has been discussed before \citep{Hubbard01, Burrows03, Guillot06, BHBH07}, but is calculated in the context of our analytical atmospheric temperature profile. I then compare the evolution of the transit radii for the different boundary conditions to the measured one.  

\subsection{Vertical optical depth \& photospheric radius}
Our approximation of a planar atmosphere is equivalent to assuming that the pressure scale height in the atmosphere $H\equiv -dr/d\ln P$ is infinitely small compared to the planetary radius, i.e. $H/r \ll 1$. In the case of HD209458b, assuming a perfect gas, a mean molecular weight $\mu=2.3$ a mean temperature $T=1500$\,K and gravity $g=980\,\rm cm\,s^{-2}$, $H={\cal R}T/\mu g \approx 550$\,km, for a planetary radius $R=94370\,$km. Therefore $H/r\approx 6\times 10^{-3}$ which is very small compared to other sources of uncertainties and justifies the planar approximation. I will therefore consider that $g$ is constant in the atmosphere.

\begin{figure}
\centerline{\resizebox{8cm}{!}{\includegraphics{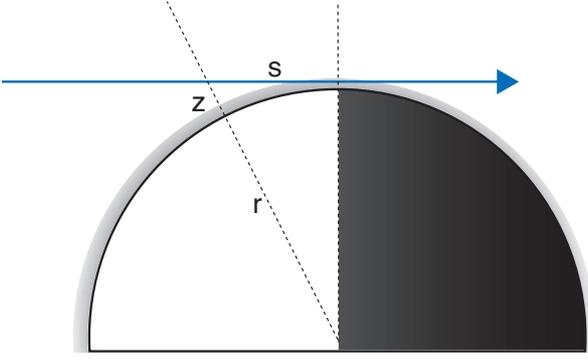}}}
\caption{Geometry of the problem for the calculation of the chord optical depth.}
\label{fig:chord}
\end{figure}

In what follows, I will use the following notation: $X(r)$ will denote a quantity that is evaluated at level $r$ but that is assumed constant in the atmosphere. The independent length variable in the atmosphere will be denoted $z$, $z=0$ corresponding to a radius level $r$ as measured from the planet's center (see fig.~\ref{fig:chord}). 

By definition of $\tau$
\begin{equation}
d\tau=-\rho \kapth dz
\label{eq:dtau}
\end{equation}
In the context of a plane-parallel, hydrostatic atmosphere, this can be integrated to show that 
\begin{equation}
\tau={\kapth\over g(r)}P,
\end{equation}
or equivalently to relate optical depth and altitude:
\begin{equation}
{d\tau\over \tau}=-{dz\over H}.
\end{equation}
Note that $H$ is a function of $g(r)$ (assumed constant in the atmosphere), but also for $T(z)$ which can vary significantly. The above equation can be integrated to yield
\begin{equation}
z=-H(r)\tht(r,z)\ln[\tau(r+z)/\tau(r)]
\label{eq:z(tau)}
\end{equation}
where $\tht(r,z)$ is a non-isothermal correction that is equal to
\begin{equation}
\tht(r,z)={1\over \ln [\tau(r+z)/\tau(r)]}{1\over T(r)}\int_{\tau(r)}^{\tau(r+z)}T(\tau')d\tau'.
\end{equation}
In the limit of an isothermal atmosphere, $\tht(r,z)=1$. In all cases, eq.~(\ref{eq:z(tau)}) may be used to evaluate the height difference between a calculated level and e.g. the photospheric level $\tau=2/3$.

\subsection{Chord optical depth \& transit radius }

When measuring the size of an exoplanet from a primary transit, the level that is probed is higher than the photospheric level. It corresponds instead to the level at which optical rays that are grazing, at the terminator, have an optical depth close to unity \citep{Hubbard01}. Neglecting refraction, we thus define a chord optical depth for this grazing incident radiation:
\begin{equation}
\tauch(\nu,r)=\int_{-\infty}^{+\infty} \rho \kappa_\nu ds
\end{equation}
As shown by fig.~\ref{fig:chord}, $(r+z)^2=r^2+s^2$, hence
\begin{equation}
\tauch(\nu,r)=2\int_{0}^{\infty} \rho \kappa_\nu {z+r \over (z^2+2rz)^{1/2}} dz.
\end{equation}
This equation may be simplified with our plane-parallel assumption ($z/r\ll 1$). We further use eqs.~(\ref{eq:dtau}) and (\ref{eq:z(tau)}) to yield
\begin{equation}
\tauch(\nu,r)={\kappa_\nu\over \kapth}\left(2r\over H(r)\right)^{1/2} \int_{0}^{\tau(r)} {d\tau\over \left[-\tht(r,z) \ln\left({\tau(r+z)\over\tau(r)}\right)\right]^{1/2}}
\end{equation}
With a new change of variable $Z=-\ln[\tau(r+z)/\tau(r)]$, we get 
\begin{equation}
\tauch(\nu,r)=\tau(r) {\kappa_\nu\over \kapth}\left({2\pi r\over H(r)}\right)^{1/2}\int_{0}^{\infty} {e^{-Z}\over \left[\pi\tht(r,Z)Z\right]^{1/2}} dZ.
\label{eq:tauchord}
\end{equation}
Using eq.~(\ref{eq:z(tau)}), we now rewrite eq.~(\ref{eq:tauchord}) at level $r+\Delta z$:
\begin{eqnarray}
\tauch(\nu,r+\Delta z)&=&\tau(r) e^{-\Delta z/H(r)\tht(r,\Delta z)} {\kappa_\nu\over \kapth}\left({2\pi r\over H(r)}{T(r)\over T(r+\Delta z)}\right)^{1/2}\nonumber\\
&&\times\int_{0}^{\infty} {e^{-Z}\over \left[\pi\tht(r+\Delta z,Z)Z\right]^{1/2}} dZ.
\end{eqnarray}
The height difference between the photospheric level for which $\tau(r)=2/3$ and the transit radius for which $\tauch(\nu,r+\Delta z)=2/3$ is:
\begin{eqnarray}
\Delta z=H(r)\tht(r,\Delta z) &&\!\!\!\ln \left\{{\kappa_\nu\over \kapth} \left({2\pi r\over H(r)}{T(r)\over T(r+\Delta z)}\right)^{1/2}\right.\nonumber\\
&&\ \left.\times\int_{0}^{\infty} {e^{-Z}\over \left[\pi\tht(r+\Delta z,Z)Z\right]^{1/2}} dZ\right\}
\label{eq:deltaz}
\end{eqnarray}
In the limit of an isothermal atmosphere, $\tht(r,Z)=1$, the integral is equal to 1 (the erf function evaluated at infinity) and the expression of the chord optical depth and height difference reduce to the relations proposed by \citet{BHBH07}. Thus in the isothermal case,
\[
\Delta z= H(r)\ln\left\{{\kappa_\nu\over \kapth} \left({2\pi r\over H(r)}\right)^{1/2}\right\}.\nonumber
\]
 In the more general case of a variable atmospheric temperature profile, eq.~(\ref{eq:deltaz}) may be easily resolved by iterations.

\subsection{Thermal evolution and sizes of transiting exoplanets}

\begin{figure}
\centerline{\resizebox{8cm}{!}{\includegraphics{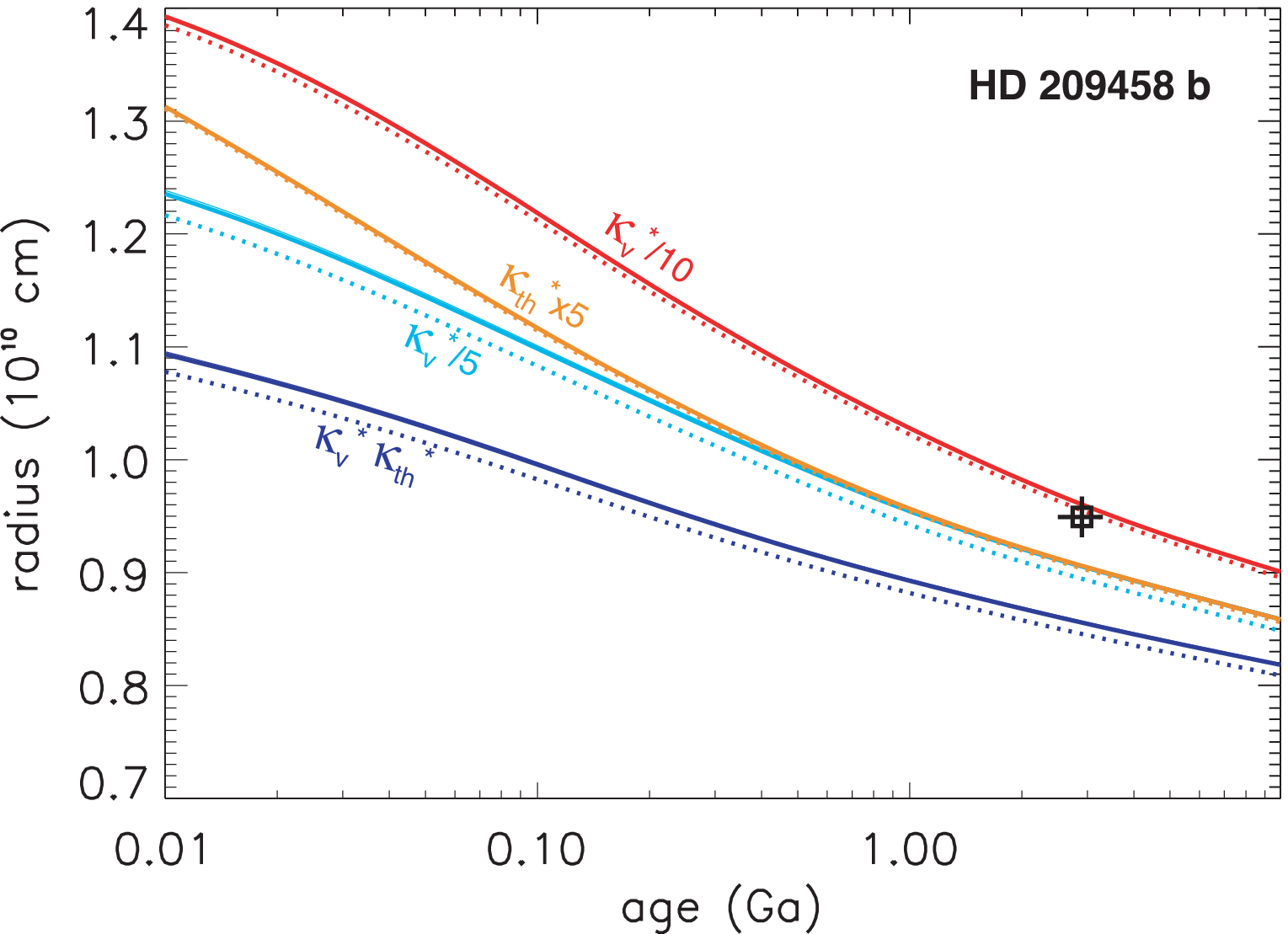}}}
\caption{Radius of HD~209458b as a function of age for different values of the thermal and visible opacities. The fiducial values are $\kapth^*=10^{-2}\rm\,cm^2\,g^{-1}$, $\kapv^*=6\times 10^{-3}\rm\,cm^2\,g^{-1}$ (bottom curves). Alternative models are found by choosing $\kapv=\kapv^*/5$, $\kapth=\kapth^*/5$, $\kapv=\kapv^*/10$, respectively. The dotted curves correspond to photospheric radii ($\tau=2/3$). The plain curves correspond to transit radii ($\tau_{\rm chord}=2/3$). The measured age and radius of the planet \citep{Knutson07b} are indicated with their error bar.}
\label{fig:evol_cepam}
\end{figure}

I now calculate how the transit radius of an exoplanet is affected by the outer boundary conditions, and specifically, using eq.~(\ref{eq:t4-global}), by a choice of $\kapth$ and $\kapv$. Figure~\ref{fig:evol_cepam} shows the result of the calculation applied to HD~209458b, assuming a low helium abundance $Y=0.24$, standard evolution models \citep{GM95,Guillot08} and with values of the opacities that vary so that $\gamma=\kappa_\nu /\kapth$ ranges between $0.04$ and $0.4$. The difference between the model radius (at 10 bars) and the transit radius in the visible is calculated using eqs.~(\ref{eq:z(tau)}) and (\ref{eq:deltaz}). With our fiducial values of the opacity coefficients, the modeled size falls short of the observed value by more than 10\%, as obtained before \citep{Bodenheimer01,GS02,Burrows03,Baraffe03,Guillot08,MFJ09}. Accounting properly for the transit radius and not the photospheric radius has a relatively small effect in that case. Decreasing the value of $\gamma$ (either by increasing $\kapth$ or by decreasing $\kapv$) does help in reducing the mismatch, but an order of magnitude increase is required in order to reproduce the observed value (even though this assumes a low helium abundance and no central core).

Clearly, this order of magnitude change of the opacity ratio compared to the fiducial value is well outside the range of possibilities spanned by present-day models: As shown in fig.~\ref{fig:compare_4pi}, the corresponding solution (dotted curve) is 700 to 1000\,K hotter than elaborate radiative transfer models predict. This remains true even when considering only the dayside average (fig.~\ref{fig:compare_2pi}). It therefore appears that alternative solutions involving additional sources of energy or non-radiative downward energy transport are required to explain the inflated sizes of exoplanets.

\section{Limitations \& complications}\label{sec:limitations}

\subsection{General approximations}

It should be stressed that, as the Eddington relation, the relations that were derived are only approximate because they assume a given dependence of the radiation field with direction (its quasi-isotropy). The discrepancy with the exact solutions should however be small \citep[see][]{Chandra60,Mihalas78} compared to the other sources of uncertainties of the problem. 

More importantly, because a grey approximation is used, this means that we cannot study the consequences of wavelength-dependent absorption. As discussed previously, the presence of strong absorption lines will lead to an efficient absorption of the energy in the higher atmosphere while allowing a deeper penetration in the wings of these lines. This cannot be captured within the framework of the simple model presented here. 

\modif{Using a plane-parallel approximation implies that solutions at very low inclination angles are probably crude compared to the true spherical geometry. This generally tends to yield an overestimation of the magnitude of the temperature inversion for low inclinations.}

Finally, departures from local thermodynamic equilibrium may be significant at relatively high altitudes and are ignored in the present study. This should be relatively minor however when concerned with the temperature profile at relatively deep levels.

\subsection{Convection}

In this work, I have considered only purely radiative solutions, with however the possibility of horizontal advection (i.e. advection along surfaces of constant pressure and therefore constant optical depth). The comparison with available models of HD~209458b has shown that it is indeed a good approximation to present models of the atmosphere of this planet and in general this should remain true for all heavily irradiated planets as these are bound to possess thick external radiative layers \citep{Guillot96,GS02}. However, as noticed by \citet{Burkert+05} and \citet{Dobbs-Dixon08}, depending on the efficiency of horizontal advection, convection should be present at lower optical depths on the night side of these planets. In this case, the problem is modified: the presence of a non-radiative vertical transport of heat invalidates the hypotheses made in \S~\ref{sec:mean T}. This implies that at optical depths for which convection is present, the mean temperature profile should depart from that predicted by eq.~(\ref{eq:t4-global}). Because convection would occur preferentially in regions with the lowest photospheric temperatures, the effect would be a more efficient loss of the internal heat, or equivalently, a lower mean interior temperature. This would generally increase the radius problem.

\subsection{Advection \& variable opacities}

Another complication is through the likely modifications of opacities and cloud coverage with temperature and irradiation level. As proposed by \citet{SG02}, the significant temperature variations in the atmosphere coupled to the horizontal (and possibly vertical) transport should affect the chemistry of the atmosphere. This is particularly true for condensing species, which could form clouds in colder region, settle to greater depths and be present in lower-than-expected abundances on the day side. This in fact may explain the relative lack of Na observed in HD209458b \citep{SG02,Iro05}. The consequence of a variable opacity field is that, even if advection is purely horizontal (in terms of pressure), the averaging performed in \S~\ref{sec:mean T} becomes invalid, as in the case of convection. This is because we cannot consider that advection proceeds on constant $\tau$ levels. Again, the effect has been investigated \citep{Burkert+05,Dobbs-Dixon08,Dobbs-Dixon10}, but in simulations in which the interior entropy was held constant, so that the consequence in terms of heat loss or internal temperatures has not been quantified. Qualitatively however, we can notice that opacities that increase with lower temperature/lower irradiation levels do tend to allow heat to penetrate efficiently into the planet near the substellar point while suppressing its loss in low-irradiation regions. This would favor a slower cooling and would therefore tend to decrease the discrepancy between models and observations. However, we can see from fig.~\ref{fig:compare_2pi} that even if we would consider that one half of the surface of the atmosphere (the night side) is not participating in the cooling, we are still a factor $\sim 4$ short in terms of opacities to explain the observations. This problem should be investigated further but it is unlikely that it can explain the size discrepancy by itself.

\subsection{Non-conservative advection, kinetic energy transport}

Finally, it should be noted that advection is not necessarily conservative, that waves and shocks may occur, and that energy may be transported as kinetic energy instead of heat. This is a complex problem \citep[e.g.][]{Goodman09} and I only mention here that, only of order $1\%$ of the incoming stellar energy needs to be transfomed into kinetic energy and dissipated at deeper levels to modify the evolution of close-in exoplanets \citep{GS02,Guillot08}. Again, this would potentially alter the modeled atmospheric temperatures.

\section{Conclusion}

An analytic {\rmod semi-grey} model to approximate the structure of plane-parallel irradiated planetary atmospheres was derived {\rmod in the framework of the Eddington approximation} [see eq.~(\ref{eq:t4-mu})]. The model is parametrized by $\kapth$ the mean opacity at thermal wavelengths, and $\kapv$ the mean opacity at the wavelengths that characterize the incoming stellar irradiation. As in the usual grey approximation, these opacities are assumed constant, however the thermal and visible opacities may differ. The relation was shown to agree with more detailed calculations in the $\tau\sim 1$ region, both as a function of the incidence angle and as a function of the mean irradiation level. The model qualitatively explains temperature inversions as resulting from a higher opacity in the optical than at thermal wavelength leading to a partial absorption of the irradiation flux at high levels in the atmosphere. It explains the proportionality relation between the deep atmospheric temperature (e.g. at a 10 bar pressure level) and the equilibrium temperature seen in detailed atmospheric calculations (small departures from this proportionality are due to variations of the infrared to visible opacities with temperature). 

The model was extended to include variable irradiation and a horizontal advection of heat. In the case of a purely horizontal (on constant optical depth surfaces) conservative advection, it was shown that the mean flux is conserved, so that a mean equation for $T^4$ may be derived [eq. ~(\ref{eq:t4-global})]. Assuming that advection homogenizes deep levels because of the increase of the radiative cooling timescale, this relation should yield the proper boundary condition for internal structure and global evolution models. The temperature that is obtained is shown to be extremely close to the temperature obtained in a one-dimensional radiative transfer model assuming isotropy of the incoming irradiation and a mean flux $\sigma\teq^4$ that corresponds to an average over the entire planetary surface [eq. ~(\ref{eq:t4-isotropic})]. 

A comparison of the results of the analytical model and of various available radiative transfer models for the transiting planet HD~209458b shows that the deep temperatures (at pressure below about 10 bars) that are obtained are generally about $\sim 1000\,$K too low to account for the observed size of the planet. Matching the observed and modeled radii requires a tenfold increase of the ratio of the infrared to the visible opacity in the atmosphere. This appears to be unlikely but the possibility merits to be investigated further given the ensemble of possibilities that remain in terms of atmospheric compositions and opacity sources. Alternatively, variations in the opacities (with higher thermal opacities in cold regions, possibly due to condensation) and kinetic energy transport are possible means to explain the size discrepancy by slowing the cooling of the planet. Progresses should be made by directly coupling radiative transfer calculations to global circulation models.

\section*{Acknowledgements}
I thank J. Fortney and T. Barman for discussions on the characteristics of planetary atmospheres, the CNRS program {\it Origine des Plan\`etes et de la Vie} and the {\it Programme National de Plan\'etologie} for support. 

\bibliography{atmospheres.bib}

\modif
\section*{Appendix: An alternative derivation}

The temperature-optical depth relation described by eq.~(\ref{eq:t4-mu}) was derived using two Eddington coefficients, $\fKth\equiv \Kth/\Jth=1/3$ and $\fHth\equiv \Hth(0)/\Jth(0)=1/2$ equivalent to the assumption that the thermal flux remains isotropic, even at low optical depth. Another derivation is possible by imposing that the emergent flux should be equal to the sum of the intrinsic and irradiated fluxes \citep{Mihalas78,Hansen08}. In this case, one still assumes $\fKth=1/3$, but instead of using $\Jth(0)=\Hth(0)/\fHth(0)$, we directly solve for $\Jth(0)$ using eq.~(\ref{eq:Jth final}) and a relation obtained from integrating the equation of radiative transfer over thermal wavelengths:
\begin{equation}
\Ith(\tau=0,\mu_*,\mu)={1\over \mu}\int_0^\infty B(t)e^{-t/\mu}dt,
\end{equation}
where $\Ith(\tau=0,\mu_*,\mu)$ is the intensity of thermal radiation emitted in direction $\mu$ from a point in the atmosphere which is irradiated by the star at an angle $\mu_*$. From eq.~(\ref{eq:rad eq}), we obtain that
\begin{equation}
\Ith(0,\mu_*,\mu)=\int_0^\infty \left[\Jth+\gamma \Jv\right]{e^{-t/\mu}\over \mu} dt.
\end{equation}
Using eq.~(\ref{eq:Jth final}) and integrating, one gets
\begin{eqnarray}
\Ith(0,\mu_*,\mu)&=&\Jth(0)+{H\over \fKth}\mu +\Jv(0){\mu_*^2\over \fKth\gamma} \nonumber\\
&&+ \Jv(0)\left(\gamma-{\mu_*^2\over \fKth\gamma}\right){1\over 1+\gamma\mu/\mu_*}.
\end{eqnarray}
We then impose that the flux emerging from the surface should be equal to the incoming flux, $4H+4\mu_* \Jv(0)$:
\begin{equation}
2\int_0^\infty \mu\Ith(0,\mu_*,\mu)d\mu=4H+4\mu_* \Jv(0).
\end{equation}
This allows expressing $\Jth(0)$ as a function of $H$ and $\mu_*\Jv(0)$. Using then eqs.~(\ref{eq:rad eq}) and (\ref{eq:Jth final}), one gets after some calculations:
\begin{eqnarray}
T^4&=&{3\tint^4\over 4}\left[{2\over 3}+\tau\right]+\tirr^4\mu_*\left\{1+{1\over 2}\left(3{\mu_*\over \gamma}-{\gamma\over\mu_*}\right)\right.\nonumber\\
&&\qquad\times\left.\left[{\mu_*\over \gamma} - {\mu_*^2\over\gamma^2}\ln\left(1+{\gamma\over\mu_*}\right)-{1\over 2}e^{-\gamma\tau/\mu_*}\right]\right\}.
\label{eq:t4-mu-outrad}
\end{eqnarray}
This equation is almost identical to the one derived by \citet{Hansen08}. However, a difference arises: the factor $(3\mu_*/\gamma-\gamma/\mu_*)$ has replaced Hansen's $(3\mu_*/\gamma)$. This is because the assumption of local thermal equilibrium implies that $B=\Jth+\gamma \Jv$ whereas \citet{Hansen08} assumes $B=\Jth$. Neglecting the $\gamma \Jv$ term in the calculation of the source function implies that direct heating from the irradiation flux is not considered: the atmosphere is heated only through the absorption of thermal radiation. In reality, the heating that is caused by the absorption of visible photons should be included, and it becomes a dominant source of heating when $\gamma/\mu >\sqrt{3}$.

Equation (\ref{eq:t4-mu-outrad}) is otherwise different in its form than eq.~(\ref{eq:t4-mu}), but they have very similar properties. For example, at the surface, for an infinite penetration of the visible flux ($\gamma\rightarrow 0$), then $T^4(\tau=0)\rightarrow (\tint^4+\mu_*\tirr^4)/2$, ie the atmosphere still behaves as if it was transporting a flux $\sigma\teff^4=\sigma\tint^4 +\sigma\tirr^4$ from below. 

As described in \S~\ref{sec:mean T}, eq.~(\ref{eq:t4-mu-outrad}) may be averaged to remove the dependence on $\mu_*$:
\begin{eqnarray}
\overline{T^4}&=&{3\tint^4\over 4}\left[{2\over 3}+\tau\right]+{\tirr^4\over 2} \int_0^1 \mu\left\{1+{1\over 2}\left(3{\mu\over\gamma}-{\gamma\over\mu}\right)\right. \nonumber\\
 &&\qquad\times\left.\left[{\mu\over\gamma} -{\mu^2\over\gamma^2}\ln\left(1+{\gamma\over\mu}\right) -{1\over 2}e^{-\gamma\tau/\mu}\right]\right\}d\mu.
\end{eqnarray}
Solving this integral analytically requires more work than with the simpler temperature profile. The following relations are useful:
\begin{eqnarray}
\int_0^{1/\gamma} x^2\ln\left(1+{1\over x}\right)dx&=&{1\over 3\gamma^3}\ln(1+\gamma)+{1\over 3}\ln\left(1+{1\over\gamma}\right)\nonumber\\
&&-{1\over 3\gamma}\left(1-{1\over 2\gamma}\right),\nonumber
\end{eqnarray}
\begin{eqnarray}
\int_0^{1/\gamma} x^4\ln\left(1+{1\over x}\right)dx&=&{1\over 5\gamma^5}\ln(1+\gamma)+{1\over 5}\ln\left(1+{1\over\gamma}\right) \nonumber\\
&&-{1\over 5\gamma}+{1\over 10\gamma^2}-{1\over 15\gamma^3}+{1\over 20\gamma^4}.\nonumber
\end{eqnarray}
The mean temperature profile is then shown to obey the following relation:
\begin{eqnarray}
\overline{T^4}&=&{3\tint^4\over 4}\left({2\over 3}+\tau\right)+\teq^4\left[{11\over 30}+{4\over 15}\gamma+{1\over 5\gamma}+{3\over 5\gamma^2}\right. \nonumber\\
&&\qquad+\left.\left({1\over 3\gamma}-{3\over 5\gamma^3}\right)\ln(1+\gamma)-{4\gamma^2\over 15}\ln\left(1+{1\over\gamma}\right)\right.\nonumber\\
&&\qquad+\left.{\gamma\over 2}E_2(\gamma\tau)-{3\over 2\gamma}E_4(\gamma\tau)\right]
\label{eq:t4-global-outrad}
\end{eqnarray}
Although the expression is more complex than eq.~(\ref{eq:t4-global}) which was derived using the second Eddington coefficient, the two are quantitatively very similar. 

\begin{figure}
\includegraphics[width=\hsize]{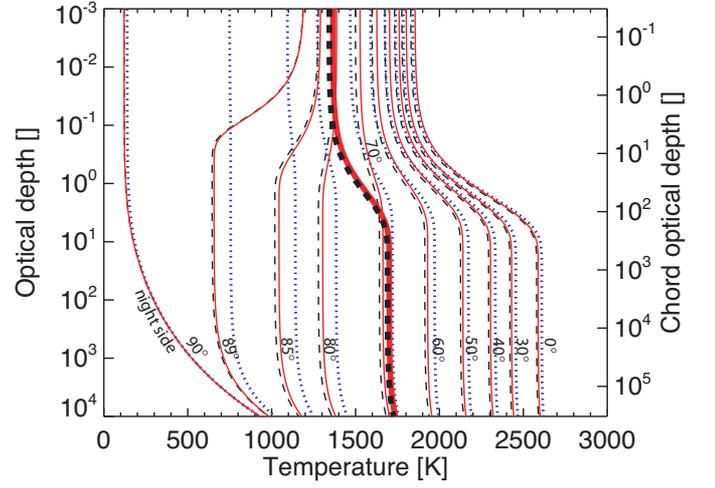}
\caption{Temperature-optical depth profiles for different inclinations of the incident light. The figure is as fig.~\ref{fig:compare_mus_tau}, but it compares the solutions obtained from eq.~(\ref{eq:t4-mu}) (dashed black lines), eq.~(\ref{eq:t4-mu-outrad}) (plain red lines), and the solutions of \citet{Hansen08} (dotted blue lines). The thick curves in the middle correspond to global averages (see eqs.~(\ref{eq:t4-global}) and (\ref{eq:t4-global-outrad}) respectively).}
\label{fig:compare_mus_tau_bis}
\end{figure}

A comparison of the various approaches is provided in fig.~\ref{fig:compare_mus_tau_bis}. Clearly, although the change in outer boundary condition affects the form and complexity of the analytical solutions, the quantitative differences are extremely small, especially when compared to the large differences between published models for exoplanets (see \S~\ref{sec:comparison}). However, there are noticeable differences with the solutions provided by \citet{Hansen08} for low $\mu$ values. In particular, temperature inversions that should occur either for low $\mu$ or high $\gamma$ values are absent of Hansen's solutions, a direct consequence of neglecting the heating caused by absorption of visible radiation. 

\end{document}